\documentclass[fleqn,twoside]{article}
\usepackage{graphicx}
\usepackage{a4,epsfig}
\usepackage{amsmath}
\usepackage{textcomp}
\usepackage{multirow}
\usepackage{espcrc2}
\usepackage[figuresright]{rotating}
\newcommand{\efficiency}{ef\mbox{}f\mbox{}iciency}
\newcommand{\fit}{f\mbox{}it}
\newcommand{\fix}{f\mbox{}ix}
\title{Limits on $\nu_{\tau}$ mass by DELPHI}
\author{D. L. Perego\address[UNIMIB]
        {On behalf of the DELPHI Collaboration, \\
        Universit\`a degli Studi di Milano-Bicocca and INFN, \\
        Piazza della Scienza 3, I$-$20126 Milan, Italy \\
	E-mail: davide.perego@mib.infn.it}}
\begin{document}

\begin{abstract}
A limit  on the  tau neutrino  mass is obtained  using all  the $Z^{0}
\rightarrow  \tau^{+} \tau^{-}$ data  collected at  LEP by  the DELPHI
detector between 1992  and 1995. In this analysis  events in which one
of the taus decays into  one charged particle, while the second $\tau$
decays into f\mbox{}ive charged pions (1-5 topology) have been used.

The  neutrino mass  is determined  from a  bidimensional \fit  ~on the
invariant mass  $m^{*}_{5 \pi}$ and on  the energy $E_{5  \pi}$ of the
f\mbox{}ive $\pi^{\pm}$ system.  The result found is $m_{\nu_{\tau}}<$
48.0 MeV$\slash c^{2}$ at 95\% conf\mbox{}idence level.
\vspace{1pc}
\end{abstract}

\maketitle

\section{Introduction}
\label{sec:section1}
The question whether  neutrinos are massive is one  of the outstanding
issues in particle physics, astrophysics and cosmology.

Among    the   possible    frameworks,  the    ``see-saw''   mechanism
\cite{gellmann} is  considered to be  the most interesting  because it
explains the  smallness of neutrino  masses by connecting them  to the
scale of new physics. In fact the see-saw mechanism assumes a neutrino
mass hierarchy  similar to that of  quarks or leptons.

On the  basis of cosmological arguments  a stable tau  neutrino with a
mass larger than a few eV$\slash c^{2}$ cannot exist; however unstable
neutrinos can be  more massive, so a direct  neutrino mass measurement
of  the  order of  a  few MeV$\slash  c^{2}$  would  suggest that  the
$\nu_{\tau}$ is unstable \cite{harari}.

The  clearest  evidence  of  neutrinos  having  mass  comes  from  the
oscillation  experiments   \cite{superkam},  even  if   this  kind  of
experiments  cannot  give  the   neutrino  mass  scale.   Only  direct
measurements can  do it and  the most recent  experiments \cite{mainz}
have \fix ed the mass of the $\nu_{e}$ to be about 2 eV$\slash c^{2}$.
However   the  results   of  the   oscillation  experiments   must  be
conf\mbox{}irmed  before concluding  that  massive unstable  neutrinos
cannot exist.

The  kinematic  properties  of  the  observed  f\mbox{}inal  state  in
hadronic  $\tau$ decays  give direct  bounds on  $m_{\nu_{\tau}}$. The
best decay channels  for a missing mass measurement  are those with  a
high  hadronic  mass,  where   the  little  energy  is  available  for
undetected  neutrinos.  This kind  of $\tau$  decays presents  a small
branching fraction  , so the data  sample generally consists  of a few
events.

High multiplicity decays $\tau^{\pm} \rightarrow 3\pi^{\pm} 2\pi^{\mp}
\overline{\nu}_{\tau}  (\nu_{\tau})$  have   been  selected  for  this
analysis.  The  end point  of the invariant  mass distribution  of the
hadronic  system,  compared  to  $m_{\tau}$,  gives  a  limit  on  the
$\nu_{\tau}$ mass.

Several  direct  experimental  measurements  have  been  done  on  the
$\nu_{\tau}$ mass.  The  world's best limit was obtained  by the ALEPH
Collaboration   \cite{aleph1,aleph2}   with   $m_{\nu_{\tau}}<$   18.2
MeV$\slash c^{2}$ at 95\% conf\mbox{}idence level.

\section{The method}
\label{sec:section2}
Multi-hadron decays of the  $\tau$ lepton \cite{okun,gomez} are due to
the couplings  of the  charged weak current  to hadrons, which  in the
case of $\tau^{\pm}  \rightarrow 5\pi^{\pm} \nu_{\tau}$ decays reduces
to the coupling of the $W^{\pm}$ boson to the $u d'$ current.

We can describe the tau decay as a two-body decay:
\begin{equation}\hspace*{-0.90cm}
	\tau^{-} \left( E_{\tau},\vec{p}_{\tau} \right) \rightarrow
	h^{-} \left( E_{h},\vec{p}_{h} \right) + \nu_{\tau} \left(
	E_{\nu},\vec{p}_{\nu} \right)
\label{eq:cinematica1}
\end{equation}
where the hadronic system $h^{-}$ is composed by f\mbox{}ive pions. In
the  tau rest  frame, the  energy of  the hadronic  system is  used to
compute  the value of  $m_{\nu_{\tau}}$ because  of its  dependence on
$m_{\tau}$, $m_{h}$ and $m_{\nu_{\tau}}$:
\begin{equation}\hspace*{-0.90cm}
	E^{*}_{h} = \frac{m^{2}_{\tau} + m^{2}_{h} - m^{2}_{\nu_{\tau
	}}}{2 m_{\tau}}
\label{eq:cinematica2}
\end{equation}
In the laboratory frame this energy becomes:
\begin{equation}\hspace*{-0.90cm}
	E_{h} = \gamma \left( E^{*}_{h} + \beta p^{*}_{h} \cos{\theta}
	\right)
\label{eq:cinematica3}
\end{equation}
where $\beta$  is the tau velocity,  $\gamma$ is the  boost factor and
$\theta$ is the angle between the  direction of the $\tau$ and that of
the hadronic system in the tau  rest frame.  We assume that the $\tau$
energy, $E_{\tau}$, is equal  to the beam energy, $E_{beam}$.  Initial
(ISR) and f\mbox{}inal (FSR) state radiation can reduce $E_{\tau}$ and
will be considered later in section \ref{sec:section6}.

\begin{figure}[!t]
\includegraphics[angle=0, width=0.50\textwidth]{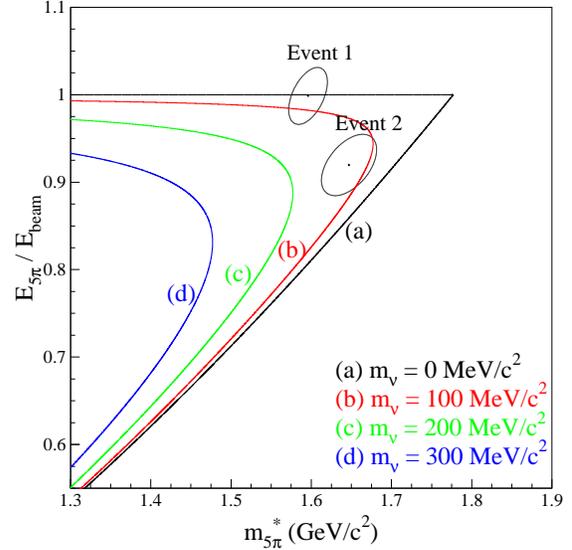}
\caption[Kinematic   allowed  regions]{Two   hypothetical  $\tau^{\pm}
\rightarrow  5\pi^{\pm} \nu_{\tau}$  events have  been plotted  in the
$\left( m^{*}_{5  \pi}, E_{5 \pi} \slash E_{beam}  \right)$ plane with
their error  ellipses. The coloured lines bound  the kinematic allowed
regions for dif\mbox{}f\mbox{}erent values of $m_{\nu_{\tau}}$.}
\label{fig:cinematico}
\end{figure}

Because of the undetected neutrino, the direction of the $\tau$ is not
determined, so $m_{\nu_{\tau}}$  cannot be computed directly. However,
the  energy of  the hadronic  system is  helpful because  it partially
recovers  the  loss  of   information.  In  fact  $E_{h}$  depends  on
$\cos{\theta}$ and it must fall inside the interval $E_{min} \le E_{h}
\le E_{max}$:
\begin{equation}\hspace*{-0.90cm}
	E^{min, max}_{h} = \gamma \left( E^{*}_{h} \pm \beta
	p^{*}_{h} \right)
\label{eq:cinematica4}
\end{equation}
As  a consequence  kinematic  allowed regions  are def\mbox{}ined  for
dif\mbox{}ferent  values  assumed  by  $m_{\nu_{\tau}}$, as  shown  in
Fig. \ref{fig:cinematico}.

The probability to f\mbox{}ind an  event in the allowed region clearly
depends on the kinematic of  the decay and on the resonance structure.
Because both  the invariant  mass $m^{*}_{5 \pi}$  and the  $E_{5 \pi}
\slash E_{beam}$ of the hadronic  system are functions of the measured
momenta, a  positive correlation arises between  these two quantities.
The error ellipses for each event must be taken into account, so that,
as  shown  in  Fig.   \ref{fig:cinematico},  event  1  constrains  the
neutrino mass  $m_{\nu_{\tau}}$ much  more than event  2, even  if the
latter  has  a  higher  hadronic  mass.   This  is  the  advantage  of
f\mbox{}itting the  distribution of the two  variables $m^{*}_{5 \pi}$
and $E_{5 \pi} \slash E_{beam}$, rather than $m^{*}_{5 \pi}$ alone.

\section{The DELPHI detector}
\label{sec:section3}
The DELPHI  detector and its  performances are described in  detail in
\cite{DELPHIde1,DELPHIde2}.   The whole detector,  composed by  a main
central   part   (\emph{barrel})    and   by   two   forward   regions
(\emph{end-caps}),  covers the  full solid  angle.   A superconducting
solenoid  provides a uniform  axial magnetic  f\mbox{}ield of  1.23 T.
Charged particles are tracked in the barrel region using a combination
of four  cylindrical subdetectors:  the silicon Vertex  Detector (VD),
the Inner  Detector (ID),  the Time Projection  Chamber (TPC)  and the
Outer  Detector  (OD).  Electron  and  photon idenif\mbox{}ication  is
provided by the High  density Projection Chamber (HPC) located outside
the OD.

\section{Data selection}
\label{sec:section4}
For  LEP  centre-of-mass  energies  of  $\sqrt{s}  \simeq  M_{Z^{0}}$,
$\tau^{+}$   and  $\tau^{-}$   are  produced   back-to-back  (ignoring
radiative ef\mbox{}fects).  Each $\tau$ decays producing one, three or
more  charged particles  in  addition  to one  or  two neutrinos  and,
possibly, neutral  mesons.  All particles apart from  neutrinos can be
detected  by  DELPHI.   $e^{+}  e^{-}  \rightarrow  Z^{0}  \rightarrow
\tau^{+}  \tau^{-}$  decays  are  easy  to identify  because  the  two
back-to-back jets  are narrow and with low  charged multiplicity.  The
presence of neutrinos implies that not  all the energy in the event is
seen  and  the invariant  mass  of  the  f\mbox{}inal state  particles
$m^{*}_{h}$ is less than $m_{\tau}$.

The background  from multi-hadron production  $e^{+} e^{-} \rightarrow
Z^{0}  \rightarrow  q \overline{q}$  at  LEP  energies  can easily  be
reduced  since  it  is  characterized  by a  relatively  high  charged
multiplicity and  by a  high  invariant mass. Final states from $e^{+}
e^{-} \rightarrow  e^{+} e^{-}$  and $e^{+} e^{-}  \rightarrow \mu^{+}
\mu^{-}$  leave  very  characteristic  signatures in  DELPHI  and  are
ef\mbox{}fectively discriminated.

The analysis presented  here is based on the  data collected by DELPHI
from 1992 to 1995; the  centre-of-mass energy $\sqrt{s}$ of the $e^{+}
e^{-}$ system was between 89 and 93 GeV.

\subsection{$\boldsymbol{e^{+}  e^{-}  \rightarrow  Z^{0}  \rightarrow
\tau^{+} \tau^{-}}$ selection}
\label{sub:tautau}
The f\mbox{}irst step in data  selection was to reject all non$-Z^{0}$
decay (cosmics, secondary interactions in the vacuum pipe, etc.). Each
event  has been  divided into  two hemispheres  def\mbox{}ined  by the
plane  perpendicular to  the thrust  axis.  It was  required that  the
reconstructed  topology  was one  charged  particle  in  one side  and
f\mbox{}ive in the other (1-5 topology).

Events were accepted  if $P_{vis}$, the sum of the  momenta of all the
charged particles  plus the total neutral  electromagnetic energy, was
greater  than 0.09$\cdot  \sqrt{s}$,  rejecting some  of the  hadronic
$Z^{0}$   decays,   $\gamma   \gamma$   interactions  and   beam   gas
interactions.   To  reject   cosmics   and  to   reduce  further   the
contamination   from   beam  gas   events   it   was  required   that,
def\mbox{}ining  particles 1  and  2  as the  most  energetic in  each
hemisphere, $r_{1,2}  <$ 1.5  cm and $\vert  z_{1,2} \vert <$  4.5 cm,
where  $r_{j}$  and  $z_{j}$  are their  transverse  and  longitudinal
distances of closest approach to the average beam spot.

Part of the  hadronic $q \overline{q}$ background was  rejected by the
requirement on the topology and  on the impact parameters as described
above.  The  remaining low  multiplicity $q \overline{q}$  and $\gamma
\gamma$ events can be easily  rejected by requiring that the isolation
angle  $\theta_{iso}$ be greater  than 160\textdegree.   The isolation
angle  is def\mbox{}ined  as the  minimum  angle between  any pair  of
tracks belonging to opposite hemispheres.
\begin{table}[Ht]
\caption[Data Sample]{Composition of the selected data sample}
\label{tab:data}
\newcommand{\m}{\hphantom{$-$}}
\newcommand{\cc}[1]{\multicolumn{1}{c}{#1}}
\renewcommand{\tabcolsep}{0.4pc} % enlarge column spacing
\renewcommand{\arraystretch}{1.2} % enlarge line spacing
\begin{tabular}{@{}ccc}
\hline
Background Class & \multicolumn{2}{c}{Background Source} \\
\hline
                         & 5$\pi^{\pm} \pi^{0}$  & 8.3\% \\
$\tau \rightarrow X$ (*) & 3$\pi^{\pm}$          & 2.3\% \\
                         & 3$\pi^{\pm} n\pi^{0}$ & 3.4\% \\
\hline
non$- \tau$ & $q \overline{q}$ & all at $m^{*}_{had}>m_{\tau}$ \\
\hline
\end{tabular}\\[2pt]
(*) in the f\mbox{}it region it is reasonably f\mbox{}lat
\end{table}

Another discriminating variable was  the total momentum of the 5-prong
system, $P_{5}$.  It  must be greater than 30  GeV$\slash c$.  Even if
either $\theta_{iso}$ and $P_{5}$ are related to the invariant mass of
the hadronic system, the distribution of $m^{*}_{5 \pi}$ in the region
of the end point is not af\mbox{}fected by these selections.

To  reduce further  $q \overline{q}$  it was  required that  the total
number of neutrals must be  less than six. The hadronic $Z^{0}$ decays
usually have  a large number of neutrals,  whereas $\tau^{+} \tau^{-}$
events should have no or few neutrals.
Four-fermion  background  was suppressed  by  the  requirement on  the
topology.

\begin{figure*}[Ht]
\includegraphics[angle=0, width=0.50\textwidth]{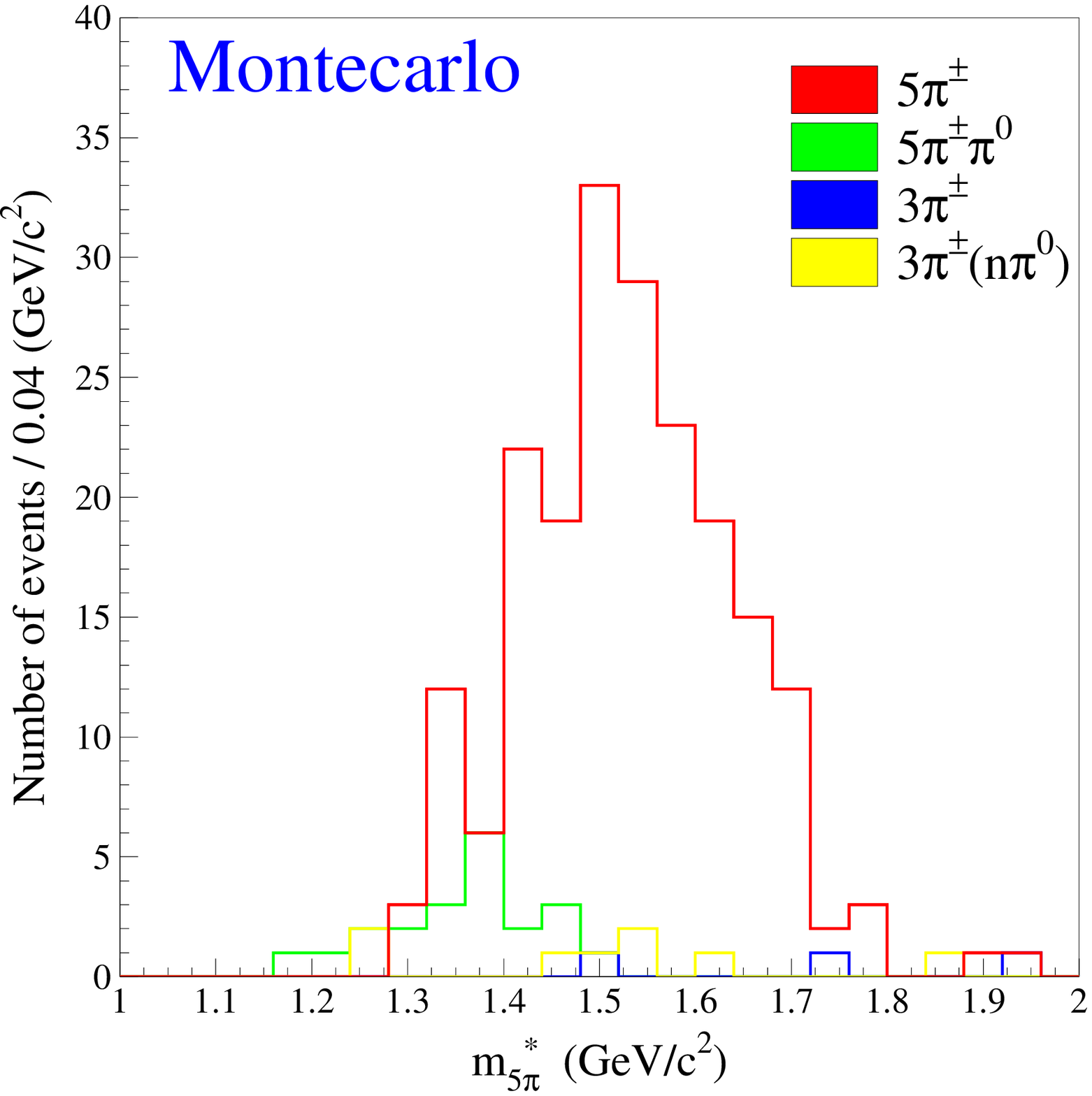}
\includegraphics[angle=0, width=0.50\textwidth]{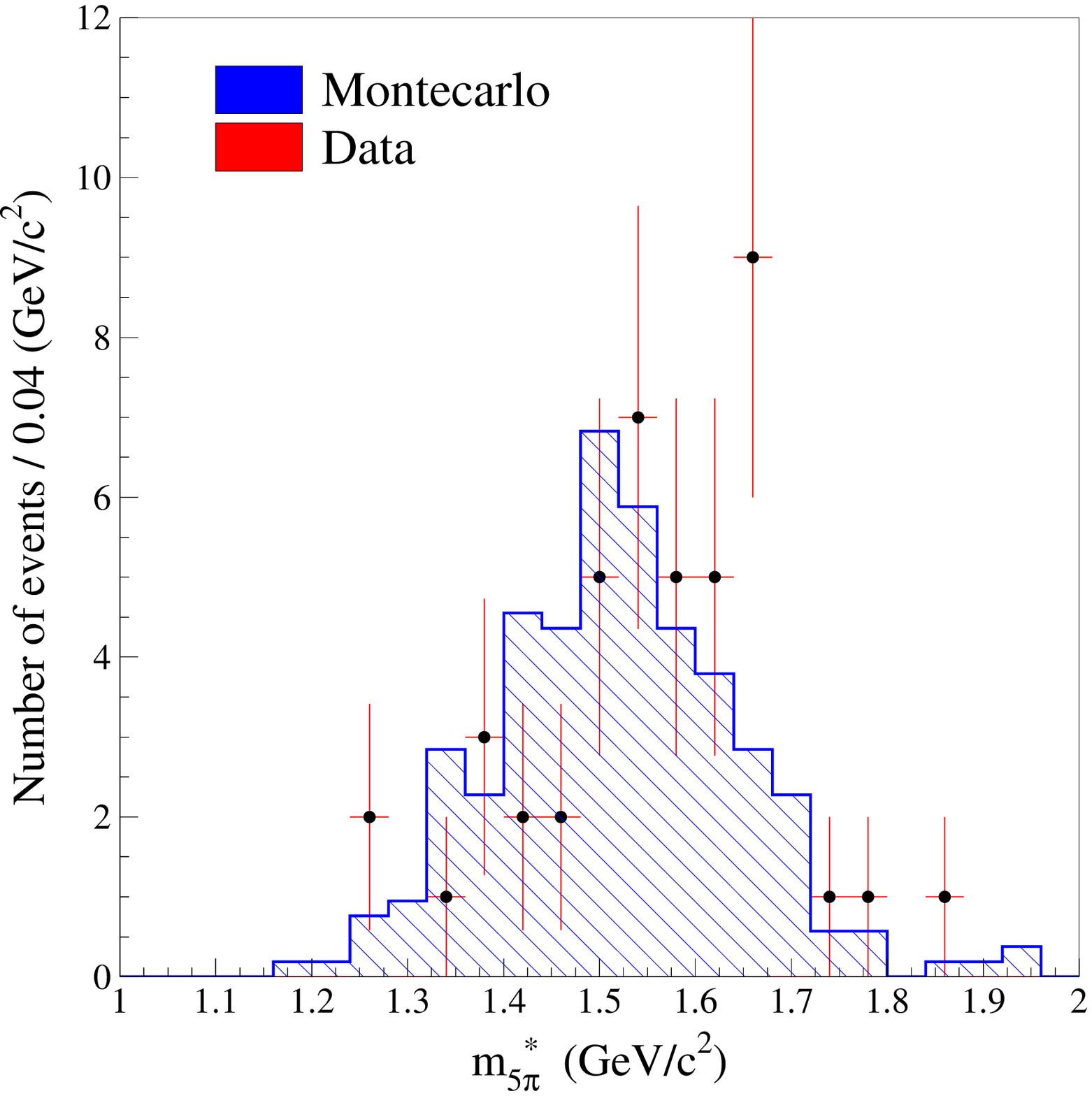}
\caption[Invariant  mass distributions]{Distribution  of  the hadronic
invariant  mass  from  simulation  (\emph{left})  and  the  comparison
between Montecarlo and data (\emph{right}).}
\label{fig:massainv}
\end{figure*}
\subsection{$\boldsymbol{\tau^{\pm}  \rightarrow 3\pi^{\pm} 2\pi^{\mp}
\overline{\nu}_{\tau} (\nu_{\tau})}$ selection}
\label{sub:channel}
The minimum  momentum among the f\mbox{}ive charged  particles and the
minimum number  of hits  in the Vertex  Detector help to  reject those
events that are reconstructed with a 5-prong multiplicity because of a
secondary   interaction  or   a  $\pi^{0}$   Dalitz   decay  ($\pi^{0}
\rightarrow   \gamma  e^{+}  e^{-}$),   as  for   example  $\tau^{\pm}
\rightarrow  3\pi^{\pm}  \nu_{\tau}\ge\pi^{0}$.    For  this  kind  of
events,   secondary  interactions   or  successive   decays   cause  a
degradation of the outcoming  momenta. Events with $P_{min}$ less than
0.5 GeV$\slash c$ were rejected.

The minimum invariant mass of pairs of oppositely charged particles in
the hemisphere  containing f\mbox{}ive tracks,  $m^{*}_{ee}$ (assuming
the  electron mass for  both), was  required to  be greater  than 0.05
GeV$\slash c^{2}$,  to reject events with a  $\gamma \rightarrow e^{+}
e^{-}$    conversion.    The   $\tau^{\pm}    \rightarrow   5\pi^{\pm}
\nu_{\tau}\ge \pi^{0}$ decays are rejected requiring the maximum value
of the electromagnetic energy being less than 4 GeV.

\subsection{Final data sample}
\label{sub:final}
The composition  of the selected  data sample, as determined  with the
Montecarlo, is listed in Table \ref{tab:data}.

A   total  of  47   $\tau^{\pm}  \rightarrow   5\pi^{\pm}  \nu_{\tau}$
candidates  have been  selected in  the data.  Fig. \ref{fig:massainv}
shows the distribution of  the hadronic invariant mass $m^{*}_{5 \pi}$
for the  simulation (signal and backgrounds) and  a comparison between
the simulation  and data.   $Z^{0} \rightarrow q  \overline{q}$ events
have values  of the invariant  mass greater than 2  GeV$\slash c^{2}$;
other kind of  $\tau$ decays have small values  of $m^{*}_{5 \pi}$, so
they are located on the left side of the invariant mass spectrum.

\section{The likelihood function}
\label{sec:section5}
In order to  determine the upper limit on $m_{\nu_{\tau}}$ an unbinned
likelihood \fit  ~has been performed. The  likelihood function depends
both on  the invariant  mass $m^{*}_{5 \pi}$  and on the  energy $E_{5
\pi}$ of the hadronic system:
\begin{equation}\hspace*{-0.90cm}
		\mathcal{L} = \prod^{N_{obs}}_{i=1} P \left(
		m^{*}_{i}, E_{i} \vert m_{\nu_{\tau}} \right)
\label{eq:likelihood1}
\end{equation}
where $N_{obs}$  is the  number of the  selected candidates,  while $P
\left(   m^{*}_{i},  E_{i}  \vert   m_{\nu_{\tau}}  \right)$   is  the
probability  for observing  each selected  event $i$  at  the position
($m^{*}_{i}  \equiv m^{*}_{5  \pi}$,  $E_{i} \equiv  E_{5 \pi}  \slash
E_{beam}$) in the kinematic plane  def\mbox{}ined as a function of the
tau neutrino mass $m_{\nu_{\tau}}$:
\begin{equation}\hspace*{-0.90cm}
	\begin{split} P  \left( m^{*}_{i}, E_{i}  \vert m_{\nu_{\tau}}
		\right) & = \frac{1}{N}\frac{d^{2} N \left( m^{*}_{i},
		E_{i}   \vert  m_{\nu_{\tau}}  \right)}{dm~dE}   \\  &
		\otimes  \mathcal{R} \left(  m,  E, \rho,  \sigma_{m},
		\sigma_{E}  \right) \\ &  \otimes \epsilon  \left( m,E
		\right) \\
\end{split}
\label{eq:likelihood2}
\end{equation}
where  $\mathcal{R}$ is  the experimental  resolution function  of the
detector   and  $\epsilon$   is  the   selection   \efficiency.   Both
$\mathcal{R}$  and  $\epsilon$  could  be  a function  of  either  the
invariant mass and the normalized energy.  The shape of the resolution
function $\mathcal{R}$ has been  determined from simulation. Since the
two variables are correlated, $\mathcal{R}$ is a 2D gaussian depending
on  the three  parameters $\sigma_{m}$,  $\sigma_{E}$ and  $\rho$. The
invariant mass  resolution is about 16 MeV$\slash  c^{2}$ (in average)
and  the normalized  energy resolution  is of  the order  of 10$^{-2}$
($\sigma_{E}  \simeq$ 500 MeV).   If the  resolution is  constant, the
experimental resolution function can be factorized, and the likelihood
$\mathcal{L}$  becomes  simpler.  In   this  analysis  each  event  is
considered with its proper resolution.

\begin{figure}[Ht]
\includegraphics[angle=0, width=0.50\textwidth]{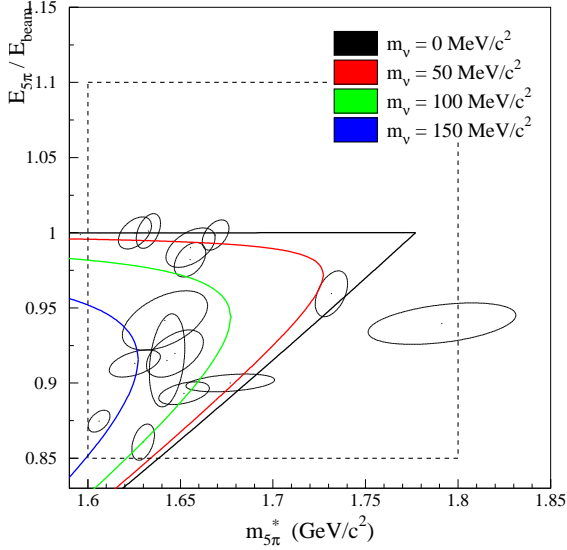}
\caption[DELPHI Collaboration]{Distribution  in the upper  part of the
$\left( m^{*}_{5  \pi}, E_{5 \pi}  \slash E_{beam} \right)$  plane for
$\tau^{\pm}   \rightarrow  5\pi^{\pm}   \nu_{\tau}$   candidates.  The
coloured   lines    bound   the   kinematic    allowed   regions   for
dif\mbox{}f\mbox{}erent values of $m_{\nu_{\tau}}$.}
\label{fig:delphimass}
\end{figure}
The selection  \efficiency ~$\epsilon$ was  found to be  constant with
respect to the invariant mass and  the energy if $m^{*}_{5 \pi} >$ 1.6
GeV$\slash  c^{2}$  and  $E_{5  \pi}  \slash  E_{beam}$  greater  than
0.85. As a consequence, the \fit ~will be restricted to this region of
the  ($m^{*}_{5  \pi}$,  $E_{5  \pi}  \slash  E_{beam}$)  plane.  Fig.
\ref{fig:delphimass} shows  this upper part  of the plane with  the 15
candidates used in the \fit.

The   theoretical  prediction   $P  \left(   m^{*}_{i},   E_{i}  \vert
m_{\nu_{\tau}}  \right)$   has  been  calculated  as   a  function  of
$m_{\nu_{\tau}}$ including initial  (ISR) and f\mbox{}inal (FSR) state
radiation with the KORALZ 4.0 \cite{koralz} Montecarlo event generator
for  $e^{+} e^{-}  \rightarrow Z^{0}  \rightarrow  \tau^{+} \tau^{-}$.
The  KORALZ 4.0  simulator incorporated  the TAUOLA  2.4 \cite{tauola}
package for modelling the $\tau$ lepton decays.

\begin{figure}[Ht]
\includegraphics[angle=0, width=0.50\textwidth]{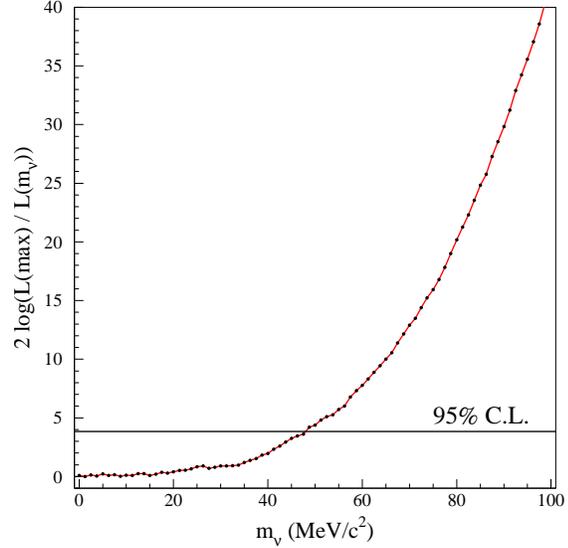}
\caption[Likelihood   of   analysis]{Log-likelihood   of   $\tau^{\pm}
\rightarrow 5\pi^{\pm} \nu_{\tau}$ data \fit ~as a function of the tau
neutrino mass. The result gives $m_{\nu_{\tau}}<$ 48 MeV$\slash c^{2}$
at 95\% CL.}
\label{fig:fit}
\end{figure}
The  $1 \slash  N$ factor  in equation  (\ref{eq:likelihood2}) ensures
that  the  probability  density   used  in  the  event  likelihood  is
normalized for any values of neutrino mass because the distribution of
events in the kinematic allowed  region of the ($m^{*}_{5 \pi}$, $E_{5
\pi} \slash E_{beam}$) plane depends on the value of $m_{\nu_{\tau}}$.

The fraction  of backgrounds has  been determined from  simulation. As
shown in  Fig. \ref{fig:massainv}, hadronic events have  values of the
invariant mass  greater than  $m_{\tau}$, while background  from $\tau
\rightarrow     3\pi^{\pm}\ge1\pi^{0}\nu_{\tau}$     or    $\tau^{\pm}
\rightarrow  5\pi^{\pm} \pi^{0}\nu_{\tau}$  gives contribution  in the
low invariant mass   part of the spectrum. The  end point of $m^{*}_{5
\pi}$ is therefore  reasonably background-free.  However a f\mbox{}lat
background (6\% of signal) has  been considered, so the likelihood was
expanded as:
\begin{equation}\hspace*{-0.90cm}
		\mathcal{L} = \alpha~\mathcal{L}_{signal} +
		( 1 - \alpha ) ~\mathcal{L}_{bgd}
\label{eq:likelihood3}
\end{equation}
where $\alpha$ is the background fraction.  Only 15 of the 47 selected
candidates are  in the  f\mbox{}it region.  The  \fit ~gives  an upper
limit on the neutrino mass of $m_{\nu_{\tau}}<$ 48.0 MeV$\slash c^{2}$
at 95\% CL.  Fig. \ref{fig:fit} shows the log-likelihood  for these 15
events.

A possible bias  in the analysis method has  been investigated by \fit
ting  high statistics Montecarlo  samples with  three dif\mbox{}ferent
input values of $m_{\nu_{\tau}}$: 0,  30 and 60 MeV$\slash c^{2}$.  In
all cases the best \fit ~results  are in good agreement with the input
values.     The   corrisponding   values    are   listed    in   Table
\ref{tab:bestfit}.
\begin{table}[Ht]
\caption[Testing the  likelihood]{Comparison between input  values for
the $\nu_{\tau}$ mass and best \fit ~results}
\label{tab:bestfit}
\newcommand{\m}{\hphantom{$-$}}
\newcommand{\cc}[1]{\multicolumn{1}{c}{#1}}
\renewcommand{\tabcolsep}{1.9pc} % enlarge column spacing
\renewcommand{\arraystretch}{1.2} % enlarge line spacing
\begin{tabular}{@{}cc}
\hline
$m_{\nu_{\tau}}$ (MeV$\slash c^{2}$) & Best Fit (MeV$\slash c^{2}$) \\
\hline
 0.0     &    7.3$^{+30.7}_{- 7.3}$ \\
30.0     &   34.0$^{+17.2}_{-32.0}$ \\
60.0     &   55.2$^{+19.0}_{-24.0}$ \\
\hline
\end{tabular}\\[2pt]
\end{table}

The expected  best limit for $m_{\nu_{\tau}}=$ 0  MeV$\slash c^{2}$ is
about 30 MeV$\slash c^{2}$.

\section{Systematic ef\mbox{}fects}
\label{sec:section6}
Four  sources of  systematic errors  have been  considered.   For each
source  a  new \fit  ~has  been  performed  after having  changed  the
parameter  of that particular  source by  one standard  deviation. The
dif\mbox{}ference between the 95\% CL  upper limit on the tau neutrino
mass computed  in section \ref{sec:section5}  and the one  obtained by
the modif\mbox{}ied  likelihood has been considered  as the systematic
ef\mbox{}fect due to that particular source. Then all these variations
have been summed in quadrature  to obtain the global  systematic error
which can be added linearly to the upper limit of the original \fit.

\begin{table}[Ht]
\caption[Resonances]{95\%     CL    $m_{\nu_{\tau}}$     limits    for
dif\mbox{}f\mbox{}erent decay structures}
\label{tab:risonanze}
\newcommand{\m}{\hphantom{$-$}}
\newcommand{\cc}[1]{\multicolumn{1}{c}{#1}}
\renewcommand{\tabcolsep}{0.6pc} % enlarge column spacing
\renewcommand{\arraystretch}{1.2} % enlarge line spacing
\begin{tabular}{@{}lcc}
\hline
                        & $m_{\nu_{\tau}}$ (MeV$\slash c^{2}$) & CL \\
\hline
pure  phase space                      & $<$ 48.0    &  95\%        \\
$\tau^{\pm} \rightarrow a^{\pm}_{1} \pi^{\pm} \pi^{\mp}
\overline{\nu}_{\tau} (\nu_{\tau})$      & $<$ 47.9  &  95\%        \\
$\tau^{\pm} \rightarrow \rho 2\pi^{\pm} \pi^{\mp} \overline{\nu}_{\tau}
                           (\nu_{\tau})$ & $<$ 46.3  &  95\%        \\
$\tau^{\pm} \rightarrow 2\rho \pi^{\pm} \overline{\nu}_{\tau}
                            (\nu_{\tau})$ & $<$ 44.5  & 95\%        \\
\hline
\end{tabular}\\[2pt]
\end{table}

The systematic  sources considered in this analysis  are summarized in
the following:
\begin{itemize}
\item the tau  mass  $m_{\tau}$  and its  energy  $E_{\tau}$;
\item detector ef\mbox{}fects such as  the hadronic mass scale and the
experimental  resolution  function.  The  DELPHI hadronic  mass  scale
studied using  the $D^{0}  \rightarrow K^{-} \pi^{+}  \pi^{-} \pi^{+}$
decay shows a shif\mbox{}t of +4 MeV$\slash c^{2}$~\cite{hadronics};
\item the selection \efficiency ~and  the ef\mbox{}fect of the size of
the \fit  ted region;
\item the intermediate structure  of the decay. The hadronic invariant
mass depends  on the dynamics  of the decay.   Intermediate resonances
have the ef\mbox{}fect of shif\mbox{}ting the shape of $m^{*}_{5 \pi}$
spectrum to higher $q^{2}$ values. Some studies have been performed by
assuming  $\tau^{\pm}   \rightarrow  a^{\pm}_{1}  \pi^{\pm}  \pi^{\mp}
\overline{\nu}_{\tau}  (\nu_{\tau})$  decay,  $\tau^{\pm}  \rightarrow
\rho  2  \pi^{\pm}  \pi^{\mp} \overline{\nu}_{\tau}  (\nu_{\tau})$  or
$\tau^{\pm}   \rightarrow  2   \rho   \pi^{\pm}  \overline{\nu}_{\tau}
(\nu_{\tau})$.
\begin{figure}[!t]
\includegraphics[angle=0, width=0.49\textwidth]{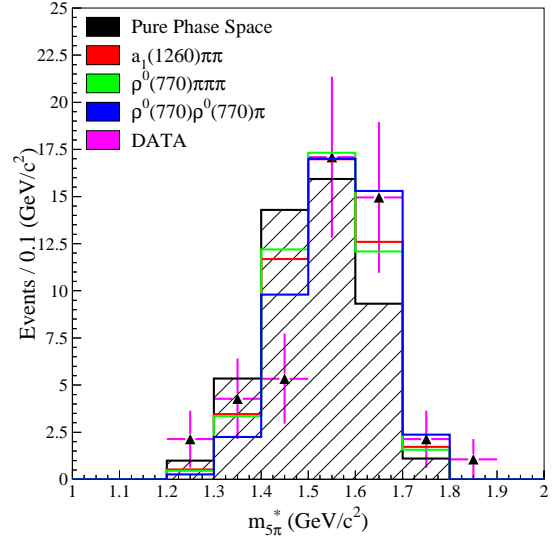}
\caption[Dynamics  of   the  decays]{Distributions  of   the  hadronic
invariant mass  for data  (dots) and four  models of decay.  The black
hatched histogram indicates the pure phase space model, while the red,
green and blue lines are  the distributions obtained by means of three
different intermediate resonance structures.}
\label{fig:risonanze}
\end{figure}
Fig.   \ref{fig:risonanze}  shows   the  invariant  mass  distribution
compared with those  predicted by a pure phase space  model and by the
models with intermediate resonances. A model with a $2 \rho \pi^{\pm}$
seems  to be  preferred.   However,  because of  the  small number  of
observed  events, it is  not possible  to state  which is  the correct
resonance.  Table  \ref{tab:risonanze} shows the 95\%  CL upper limits
obtained for the resonance structures considered.
\end{itemize}

\begin{table}[Ht]
\caption[Systematics]{Systematic variations of the 95\% CL upper limit on $m_{\nu_{\tau}}$}
\label{tab:sistematiche}
\newcommand{\m}{\hphantom{$-$}}
\newcommand{\cc}[1]{\multicolumn{1}{c}{#1}}
\renewcommand{\tabcolsep}{0.0pc} % enlarge column spacing
\renewcommand{\arraystretch}{1.2} % enlarge line spacing
\begin{tabular}{@{}cc}
\hline
\multirow{2}{50mm}{\centering Source} & Variation of $m_{\nu_{\tau}}$ \\
                        & limit (MeV$\slash c^{2}$)\\
\hline
$m_{\tau}$                  &    0.3 \\
$E_{beam}$                  & $<$0.1 \\
experimental resolution     &    2.5 \\
mass calibration            &    1.0 \\
\fit ~region                &    0.7 \\
\hline
total                       &    2.8 \\
\hline
hadronic mass scale         &   +4.0 \\
\hline
\end{tabular}\\[2pt]
\end{table}
Table \ref{tab:sistematiche} summarises  all the variations considered
in this analysis.

\section{Review of the existing collider results}
\label{sec:section7}
There are some interesting limits on the $\nu_{\tau}$ mass obtained by
collider   experiments.  Recent  results   are  summarized   in  Table
\ref{tab:ashortreview}   and  shortly   presented  in   the  following
sections.

\begin{table*}[htb]
\caption[Review]{Limits on the  $\nu_{\tau}$ mass in MeV$\slash c^{2}$
from multi-hadron $\tau$ decays}
\label{tab:ashortreview}
\newcommand{\m}{\hphantom{$-$}}
\newcommand{\cc}[1]{\multicolumn{1}{c}{#1}}
\renewcommand{\tabcolsep}{1.1pc} % enlarge column spacing
\renewcommand{\arraystretch}{1.2} % enlarge line spacing
\begin{tabular}{@{}lccccc}
\hline
\multirow{2}{43mm}{Experiment} & \multicolumn{5}{c}{Mode} \\
\cline{2-6}
	& $3\pi \nu_{\tau}$ & $3\pi \pi^{0} \nu_{\tau}$
	& $3\pi 2\pi^{0} \nu_{\tau}$ & $5 \pi (\pi^{0})
	\nu_{\tau}$ & Combined \\
\hline
ARGUS   &         &      &      & 31.0 &      \\
CLEO    &         & 28.0 & 35.9 & 33.9 & 30.0 \\
ALEPH   & 26.5    &      &      & 22.3 & 18.2 \\
OPAL    & 35.3(*) &      &      & 43.2 & 27.6 \\
\hline
\end{tabular}\\[2pt]
(*)      from     $\tau^{+}     \tau^{-}      \rightarrow     3h^{\pm}
\overline{\nu}_{\tau}+3h^{\mp} \nu_{\tau}$ decays
\end{table*}

\subsection{Low energy measurements}
\label{sub:lowenergy}
The   main  dif\mbox{}ference   between   measurements  performed   at
dif\mbox{}f\mbox{}erent  $\sqrt{s}$  values   is  that  the  selection
between high multiplicity $\tau$  decays and low multiplicity hadronic
events from  $e^{+} e^{-}  \rightarrow q \overline{q}$  becomes easier
and easier with  the increasing of the center-of-mass  energy.  So LEP
experiments     select     $\tau^{+}     \tau^{-}$     events     more
ef\mbox{}f\mbox{}iciently and with a higher purity. On the other hand,
taus are produced more abundantly at CESR.

CLEO collected tau  pairs produced at a center-of-mass  energy of 10.6
GeV.   The  analysis   was  performed  using  $\tau^{\pm}  \rightarrow
5\pi^{\pm} \nu_{\tau}$  events and \fit  ting the 36  selected events,
using the  invariant mass and the  hadronic energy. The  95\% CL upper
limit was $m_{\nu_{\tau}}<$ 33.9 MeV$\slash c^{2}$ \cite{cleo1}.

In addition CLEO  analyzed $\tau^{\pm} \rightarrow 3\pi^{\pm} 2\pi^{0}
\nu_{\tau}$ decays.  This channel can  be studied better at low energy
machine, because at LEP photons from $\pi^{0}$  conversions are highly
collimated  and  the selection more dif\mbox{}f\mbox{}icult. An  upper
limit at  the 95\% conf\mbox{}idence  level of 35.9  MeV$\slash c^{2}$
was found. The combined result found by CLEO is 30.0 MeV$\slash c^{2}$
\cite{cleo1}.

An   analysis   with   $\tau^{\pm}  \rightarrow   3\pi^{\pm}   \pi^{0}
\overline{\nu}_{\tau}      (\nu_{\tau})$      decays     was      also
done \cite{cleo2}. This  is  a  useful  decay because  of  the  larger
branching ratio  than 5-prong, and  in addition the invariant  mass is
very close to $m_{\tau}$.  From a data sample of 29058 candidates, the
obtained  95\%  CL  upper limit  on  the  tau  neutrino mass  is  28.0
MeV$\slash c^{2}$.

\begin{figure}[!t]
\includegraphics[angle=0, width=0.46\textwidth]{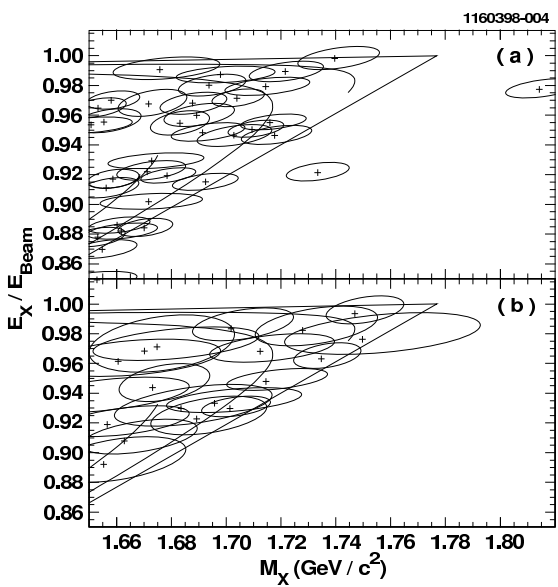}
\caption[CLEO  Collaboration]{The hadronic  scaled mass  energy versus
mass  distribution of the  $5 \pi^{\pm}$  (\emph{a}) and  $3 \pi^{\pm}
2\pi^{0}$ (\emph{b})  data sample in  the f\mbox{}it region  from CLEO
Colla\-boration.}
\label{fig:cleomass}
\end{figure}

Fig. \ref{fig:cleomass} shows the  distribution of the 5 body selected
events in the ($m^{*}_{5 \pi}$, $E_{5 \pi} \slash E_{beam}$) plane.

ARGUS   used  $\tau^{\pm}  \rightarrow   5\pi^{\pm}\nu_{\tau}$  events
collected at  center-of-mass energy of  about 10.6 GeV.   Performing a
\fit   ~only  to  the   hadronic  invariant   mass,  the   result  was
$m_{\nu_{\tau}}<$ 31.0 MeV$\slash c^{2}$ \cite{argus}.

\subsection{High energy measurements}
\label{sub:highenergy}
At  LEP  energy, $\tau^{+}  \tau^{-}$  events present  a  typical  two
back-to-back narrow  jets structure.  ALEPH and OPAL,  as DELPHI, have
analyzed data collected at center-of-mass energy equal or close to the
$Z^{0}$ resonance.

\begin{figure*}[Ht]
\includegraphics[angle=0, width=0.50\textwidth]{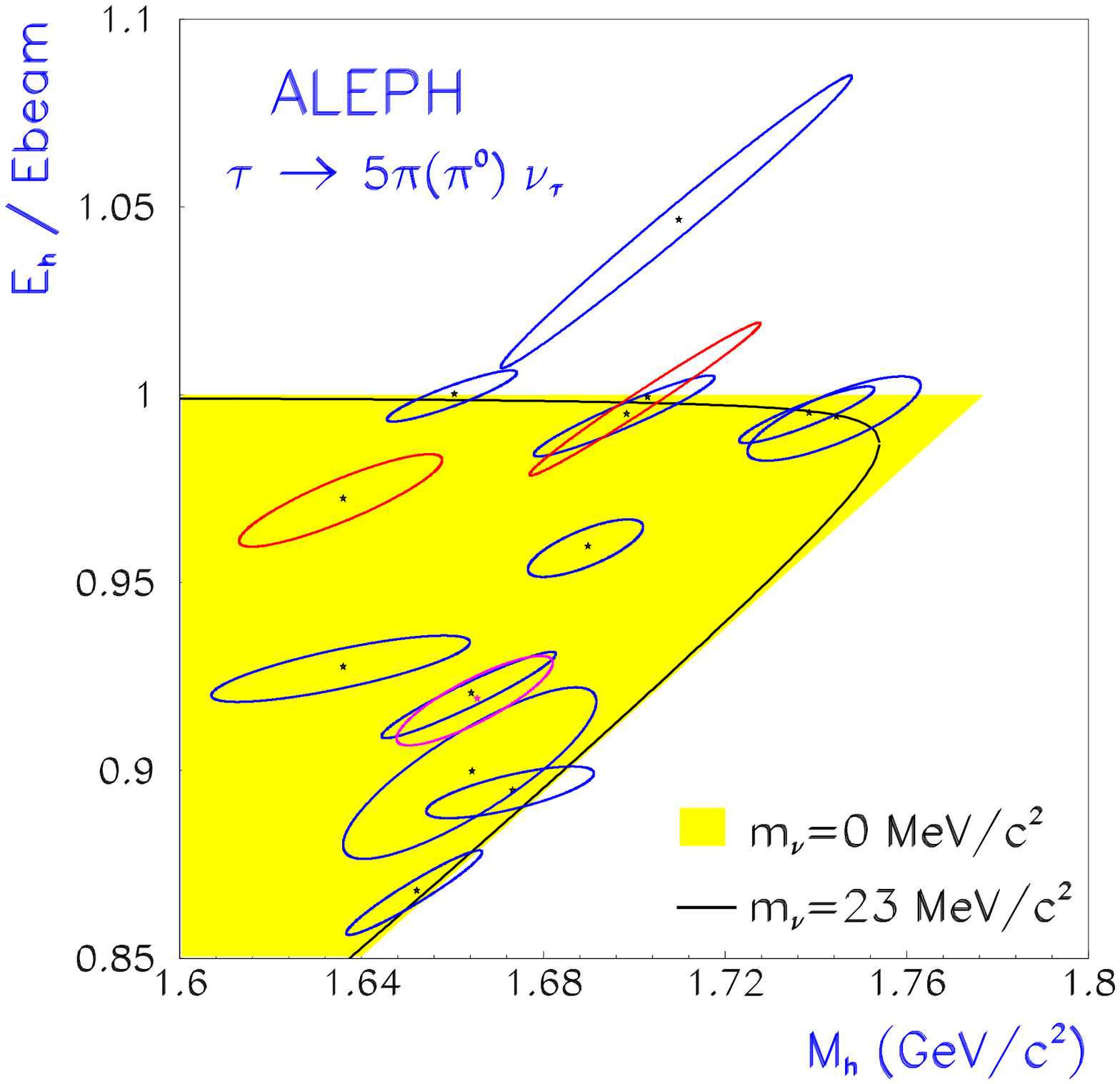}
\includegraphics[angle=0, width=0.50\textwidth]{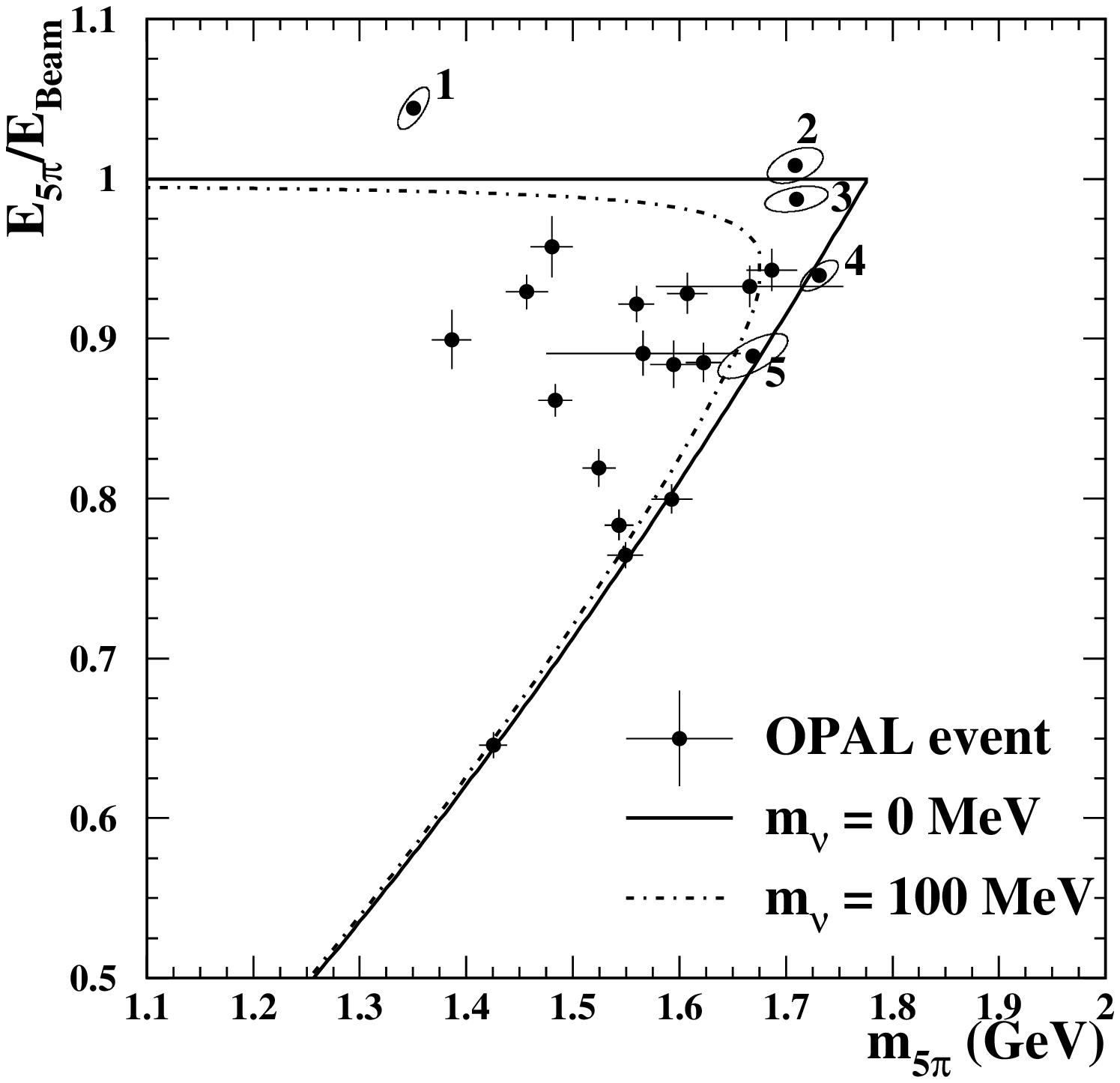}
\caption[Other LEP Experiments]{$\tau^{\pm} \rightarrow 5\pi^{\pm} \left( \pi^{0} \right) \nu_{\tau}$ candidates in the upper part of the $\left( m^{*}_{5 \pi}, E_{5  \pi} \slash  E_{beam} \right)$ plane from ALEPH (\emph{left}) and OPAL (\emph{right}).}
\label{fig:otherlepmass}
\end{figure*}

ALEPH obtained  two separate limits  by \fit ting the  distribution of
the visible  energy and  the invariant  mass. The \fit  s to  the 2939
$\tau^{\pm}   \rightarrow   3\pi^{\pm}\nu_{\tau}$   and  to   the   55
$\tau^{\pm}  \rightarrow  5\pi^{\pm} \left(\pi^{0}\right)  \nu_{\tau}$
events have  given 95\%  CL upper limits  on $m_{\nu_{\tau}}$  of 26.5
MeV$\slash c^{2}$ and 22.3  MeV$\slash c^{2}$ respectively.  These two
results have  been combined to  obtain a 95\%  conf\mbox{}idence level
upper limit  of 18.2 MeV$\slash c^{2}$,  and this is  the world's best
limit on $m_{\nu_{\tau}}$ \cite{aleph2}.  Fig.  \ref{fig:otherlepmass}
shows the distribution  of events in the upper  part of the ($m^{*}_{5
\pi}$, $E_{5 \pi} \slash E_{beam}$) plane for the 5-prong data.

Also  OPAL determined  an upper  limit for  $m_{\nu_{\tau}}$  with the
decay $\tau^{\pm} \rightarrow 5\pi^{\pm} \nu_{\tau}$.  A limit of 43.2
MeV$\slash c^{2}$ at 95\% CL has been obtained using a two-dimensional
method in the hadronic invariant mass and energy distributions from 22
selected candidates  \cite{opal1}.  Combining this  result with OPAL's
previously published measurement  using $\tau^{+} \tau^{-} \rightarrow
3h^{\pm}  \overline{\nu}_{\tau}+3h^{\mp}  \nu_{\tau}$  decays,  a  new
limit  of  $m_{\nu_{\tau}}<$  27.6  MeV$\slash c^{2}$  (95\%  CL)  was
obtained \cite{opal2}.

Before the  5-prong analysis  presented here, DELPHI  has set  a limit
with a measurement  of the leptonic branching fractions  of the $\tau$
lepton.   The  95\%  conf\mbox{}idence  level  limit  found  was  66.0
MeV$\slash c^{2}$ \cite{delphibr}.  This  method is however limited by
the precision of the leptonic branching ratio measurements, and is not
competitive with the one described in this talk.

\section{Conclusions}
\label{sec:conclusions}
The CLEO  experiment had analyzed  a very large statistics  sample and
its mass and energy resolutions were better than LEP experiments.

Looking at Fig.  \ref{fig:delphimass} and \ref{fig:otherlepmass}, it's
easy to outline that with  comparable mass and energy resolutions, the
position  of  events  in  the  ($m^{*}_{5  \pi}$,  $E_{5  \pi}  \slash
E_{beam}$)   plane  determines   a  ``\emph{luck   factor}''   for  an
experiment.   So,  the ALEPH  limit  is  strongly  driven by  the  two
candidates  lying in top  right-hand corner  of the  kinematic allowed
region. The most sensitive region of that plane for OPAL and DELPHI is
empty.  OPAL  and  DELPHI's  5-prong  analysis  present  very  similar
results.

\section*{Acknowledgemets}
\label{sec:acknowledgemets}
The  author thanks Clara  Matteuzzi, Tommaso  Tabarelli  de Fatis  and
Francisco  Matorras  for  their   helpful  suggestions  and  for  many
discussions for this analysis.

\end{document}